\documentclass[prl,twocolumn]{revtex4}

\begin{document}


\title
{\large{\bf Electromagnetic nonlinear X-waves}}
\author{C. Conti,$^1$ S. Trillo,$^{1,2}$
P. Di Trapani,$^3$ G. Valiulis,$^4$ O. Jedrkiewicz,$^3$ J. Trull$^{3,5}$}

\address{$^1$ Istituto Nazionale di Fisica della Materia (INFM)-RM3,
Via della Vasca Navale 84, 00146 Roma, Italy}

\address{$^2$ Department of Engineering,
University of Ferrara, Via Saragat 1, 44100 Ferrara, Italy}

\address{$^3$INFM and Department of Chemical, Physical and
Mathematical Sciences,
University of Insubria, Via Valleggio 11, 22100 Como, Italy}

\address{$^4$ Department of Quantum Electronics, Vilnius
University, Sauletekio al. 9, bldg. 3, LT-2040 Vilnius,
Lithuania}

\address{$^5$ Dept. Fisica i Enginyeria Nuclear, UPC Terrassa,
Spain}

\date{\today}

\begin{abstract}
Nonlinear optical media that are normally dispersive,
support a new type of localized (nondiffractive and nondispersive)
wavepackets that are X-shaped in space and time
and have slower than exponential decay.
High-intensity X-waves, unlike linear ones,
can be formed spontaneously through
a trigger mechanism of conical emission, thus
playing an important role in experiments.
\end{abstract}


\maketitle

The nonlinear response of condensed matter can
compensate for the diffractive spreading of optical beams, or the dispersive
broadening of pulses due to group-velocity dispersion (GVD), forming spatial
\cite{spatbook} or temporal solitons \cite{agrawal}, respectively.
Recent experimental results concerning
the self-focusing behavior of intense ultrashort pulses
\cite{ran96,zoz99,liu99,kopr00,eis01,tzo01} indicate, however,
that the spatial and temporal degrees of freedom cannot be treated separately.
When the three length scales
naturally associated with diffraction, GVD, and nonlinearity
become comparable, the most intriguing consequence
of space-time coupling is the possibility to form
a nondiffractive and nondispersive localized wavepacket (LWP),
namely a spatio-temporal soliton or light bullet \cite{silb90}
characterized by {\em exponentially} decaying tails.
A strict constraint for the formation of light bullets is
that the nonlinear phase changes counteract both
the linear wavefront curvature and the GVD-induced
chirp, leading to space-time focusing,
as occuring in Kerr-like focusing media
with anomalous GVD \cite{liu99,eis01}.
\newline\indent
Viceversa, a normal GVD rules out the possibility to achieve bullet-type
LWPs.  In this regime, the field evolution is known to be qualitatively
different, and involves complex phenomena such as
temporal splitting and spectral breaking \cite{ran96,zoz99,eis01}.
This is the reason why no attempts have been made to answer
the fundamental question as to {\em whether any form of nonlinearity-induced
localization could still take place in normally dispersive media}.
In this letter, we show that LWPs do exist
also with normal GVD in the form of {\em nonlinear} X-waves (NLXWs)
or X-wave solitons.
To date X-shaped waves are known only in the context
of {\em linear} acoustic \cite{acoutheory}, or electromagnetic \cite{opttheory}
propagation, and constitute the polichromatic
generalization of diffraction-free Bessel (or Durnin \cite{durnin87}) beams.
They have been observed in both acoustical \cite{lg92} and optical
\cite{saari97} experiments, both requiring beam-shaping techniques.
Here, we find propagation-invariant NLXWs that can be naturally regarded as the
continuation of linear X-waves into the nonlinear regime.
Yet, we find one {\em fundamental difference} between linear
and NLXWs. At high intensity, the formation of X-shaped LWPs
occurs spontaneously from conventional bell-shaped (in space and time) beams throughself-induced spectral reshaping triggered by a mechanism of conic
al emission.
Thus, NLXWs are expected to have stronger impact on experi\-ments
than linear X waves.
\newline\indent
To support the generality of NLXW concept,
we choose two different phenomena,
whose spatio-temporal dyna\-mics have been widely
investigated experimentally. Specifically, we consider self-action
of a scalar wavepacket $u_1$ (carrier $\omega_0$) due to
a pure (cubic) focusing Kerr effect, or generation of a $u_2$ wavepacket
at second-harmonic (SH, $2\omega_0$) in non-centrosymmetric (quadratic) media.
In the paraxial regime, the evolution in Kerr media is ruled by
the scalar 1+3 nonlinear Schr\"{o}dinger (NLS) equation,
\begin{equation}\label{nls}
i\partial_{\zeta} u_1 + \nabla_{\perp}^2 u_1 -d_1 \partial_{\tau \tau} u_1
+ \Gamma |u_1|^2 u_1 =0,
\end{equation}
whereas SH generation is ruled by the vector NLS model
\FL \begin{equation} \label{shg}
\begin{array}{l}
{\displaystyle
(i \partial_{\zeta} +
\sigma_1 \nabla_{\perp}^2
-d_1 \partial_{\tau \tau}) u_1 + \Gamma u_2 u_1^*
e^{i\delta k \zeta} =  0,} \\
{\displaystyle (i \partial_{\zeta} +
\sigma_2 \nabla_{\perp}^2 +
i v \partial_{\tau} -d_2 \partial_{\tau \tau} ) u_2 +
\Gamma \frac{u_1^2}{2} e^{-i\delta k \zeta} = 0.}
\end{array}
\end{equation}
In Eqs.~(\ref{nls}-\ref{shg}) the link with real-world
variables $X,Y,Z,T$ is as follows:
$\zeta \equiv Z/Z_{df}$ is the propagation distance in
units of diffraction length $Z_{df}=2 k_1 W_0^2$ associated with the
beam waist $W_0$,
$\nabla_{\perp}^2 \equiv \partial_{\xi}^2 + \partial_{\eta}^2$ is the
transverse
Laplacian where $(X,Y)=W_0 (\xi,\eta)$,
and $t=(T-Z/V_{g1})/T_0$ is time in a frame traveling at
group-velocity $V_{g1}$ (of $u_1$) in units of $T_0=(|k_1''| Z_{df}/2)^{1/2}$,
$k_{m}''$ ($m=1,2$) being the GVD at $m\omega_0$.
The coefficients are $d_m=k_m''/|k_1''|$, $\sigma_m=k_1/k_m$ ($\sigma_2
\simeq 1/2$),
the wavevector mismatch $\delta k=(k_2-2k_1) Z_{df}$,
and $\Gamma=Z_{df}/Z_{nl}$ (not rescaled out to quantify
the impact of nonlinearities). Here $Z_{nl}=(\chi_3 I_p)^{-1}$ in
Eq.~(\ref{nls}) and
$Z_{nl}=(\chi_2 \sqrt{I_p})^{-1}$ in Eqs.~(\ref{shg})
are nonlinear length scales associated with the
input peak intensity $I_p$ at $\omega_0$ ($|u_1|^2_{max}=1$), and
$\chi_3$ [m/W] and $\chi_2$ [W$^{-1/2}$] are standard nonlinear coefficients.
Finally $v=Z_{df} \delta V/T_0$ accounts for the walk-off
due to group-velocity mismatch (GVM) $\delta V=V_{g2}^{-1}-V_{g1}^{-1}$,
though we analyse (where not stated otherwise) the GVM-matched case $v=0$.
\newline\indent
We seek for propagation-invariant, radially-symmetric
LWPs of the form $u_1=f_1(r,t) \exp(-i \beta \zeta)$,
accompanied in Eqs.~(\ref{shg}) by a symbiotic SH
$u_2=f_2(r,t) \exp[-i(2\beta + \delta k) \zeta]$. Here
$f_{1,2}$ are real, and $\beta$ is a nonlinear phase-shift
(in Eqs.~(\ref{shg}) it makes the two waves nonlinearly phase-matched).
In the cubic case, $f_{1}$ obeys the equation
\begin{equation} \label{staz1}
\ddot{f_1} + r^{-1} \dot{f_1} -D_1 \partial_{tt} f_1
+ b f_1  + \gamma f_1^3 =0,
\end{equation}
while Eqs.~(\ref{shg}) yield
(after setting $f_{1}/\sqrt{\sigma_2} \rightarrow f_{1}$),
\begin{equation} \label{staz2}
\begin{array}{l}
{\displaystyle \ddot{f_1} + r^{-1} \dot{f_1} - D_1 \partial_{tt} f_1
+ b f_1  + \gamma f_2 f_1 =0,} \\
{\displaystyle \ddot{f_2} + r^{-1} \dot{f_2} -D_2 \partial_{tt} f_2
+ \alpha f_2  + \gamma \frac{f_1^2}{2} =0}, \nonumber
\end{array}\end{equation}
where we have set $D_m=d_m/\sigma_m$,
$\dot{f}=\partial f/ \partial r$, and
$r=|\beta| \rho \equiv (x^2+y^2)^{1/2}$
($\rho^2 \equiv \xi^2+\eta^2$), $t=|\beta| \tau$,
$\gamma=\Gamma/|\beta|$,
$b=\beta/|\beta|$ and $\alpha=(2 b + \delta k/|\beta|)/\sigma_2$
in the nondegenerate case ($\beta \neq 0$),
while $(r,t)=(\rho,\tau)$, $\alpha=\delta k/\sigma_2$, $\gamma=\Gamma$, and
$b=0$ in the degenerate case ($\beta=0$).
Eqs.~(\ref{staz1}-\ref{staz2}) must be integrated
along with the boundary conditions $f_{1,2}(r,\pm \infty)=f_{1,2}(\infty,t)=0$,
and $\dot{f}_{1,2}(0,t)=0$.
While the anomalous GVD regime ($D_m<0$) guarantees that,
for $\beta<0$ ($b,\alpha<0$), nearly separable LWPs (light bullets) exist
with exponentially decaying tails, in the normal GVD regime ($D_m>0$) the
nature
of the LWP solutions (if any) must change dramatically, because
the low-intensity exponential damping no longer takes place.
Pseudo-spectral numerical techniques (i.e., solve
Eqs.~(\ref{staz1}-\ref{staz2})
as a dynamical evolution problem in $r$ with appropriate discretization in
$t$ \cite{chqz}) are well suited to search for strongly nonseparable
objects with slow spatio-temporal decay. We have implemented such methods
(using two different algorithms), and found LWPs when $\beta \ge 0$
($b,\alpha \ge 0$). Efficient convergence occurs by employing
as a trial function (of time, at $r=0$) the real part of the waveform,
\begin{equation}\label{lin}
u_{m}=\frac{1}{\sqrt{(\Delta-i(\tau/d_m))^2+\xi^2+\eta^2}}; m=1,2,
\end{equation}
\noindent which represent X-shaped LWP solutions of
Eqs.~(\ref{nls}-\ref{shg}) in the {\em linear limit} ($\Gamma=0$).
Here $\Delta$ is a free parameter: the smaller $\Delta$,
the stronger the localization.
The LWP solutions, obtained from Eq.~(\ref{staz1})
with $\Delta=1$ in Eq.~(\ref{lin}) are shown in Fig.~\ref{fig1}.
For $b=0$ and moderate nonlinearities [$\gamma=1$, Fig.~1(a)],
the "ground-state" LWP mode has a clear X-shape
(in $x-t$, or V-shape in $r-t$ variables),
encompassing a space-time tightly confined structure,
with slow axisymmetric spatial decay ($\sim 1/r$) accompanied by
(radially increasing) temporal pulse-splitting.
In the case $b=1$, while the field maintains
its basic X-shape, it develops radial (damped) oscillations
as shown in Fig.~1(b), e.g. for the strong nonlinear case ($\gamma=10$).
\begin{figure}
\caption{Spatio-temporal field profile $u_1(x,y=0,t)$ of
NLXWs in normally dispersive ($D_1=1$) Kerr media, as obtained from
Eq.~(3) with: (a) $b=0$, $\gamma=1$; (b) $b=1$, $\gamma=10$.}
\label{fig1} \end{figure}
\begin{figure}
\caption{NLXW spatio-temporal intensity
(a) $|u_1(t,r)|^2$;  (b) $|u_2(t,r)|^2$,
obtained from Eqs.~(4) with $b=1$, $\alpha=2$, $\gamma=1$.}
\label{fig2} \end{figure}
From Eqs.~(\ref{staz2}) we obtain similar NLXWs, where the
two symbiotic LWPs $f_1$-$f_2$ can be both of ground-state type ($b=\alpha=0$),
ground-oscillatory type ($b=0, \alpha \neq 0$),
or both of oscillatory type ($b=1, \alpha \neq 0$,
with out-of-phase coherent oscillation),
an example of the latter case being shown in Fig.~\ref{fig2}.
\newline\indent
In both nonlinear processes the oscillations stem from the fact that the
spatial behavior of the low-intensity portion of the LWPs is governed by a
zero-th
order Bessel equation, which can be easily obtained by Fourier-transforming
the linear
($\gamma=0$) limit of Eq.~(\ref{staz1}) or (decoupled) Eqs.~(\ref{staz2}).
Therefore NLXWs exhibit the characteristic damped spatial oscillations of
$J_0$ Bessel functions, and can be regarded as the natural
generalization of monochromatic nondiffractive co\-nical $J_0$ beams,
whose physical realization was demonstrated
both in the linear \cite{durnin87} and nonlinear \cite{bessel} regimes.
More precisely, once the NLXW solutions $f_{m}=f_{m}(r,t)$ are
obtained,  they can be represented in the spectral domain
of transverse wavevector $K=(K_x^2+K_y^2)^{1/2}$
and frequency detuning $\Omega$, through the
Fourier-Bessel transform $S_{m}=S_{m}(K,\Omega)$, as
\FL \begin{equation} \label{spectrum}
f_{m}=
\int_{0}^{+\infty} \int_{-\infty}^{+\infty}
S_{m}(\Omega,K) J_0(K r) e^{i \Omega t} K dK d\Omega,
\end{equation}
\noindent showing their nature of a weighted
superposition of co\-nical $J_0$ beams with different frequency.
\newline\indent
Compared with their counterpart in the anomalous GVD regime
(i.e., 1+3D bullet LWP), NLXWs possess important differences:
(i) NLXWs experience weaker localization, i.e.
$1/r$ instead of exponential decay.
As a consequence the energy of X-waves~(\ref{lin}) and likely also of our NLXW
solutions,  is infinite, viz $E_1=\int_0^{a} r dr
\left[\int_{-\infty}^{\infty}dt |u_1|^2 \right]$
in Eq. (\ref{nls}) tends to infinity as the
computational  window $a \rightarrow \infty$;
(ii) unlike bullets that vanish in the
linear limit,  NLXWs have a finite limit for $\Gamma=0$;
(iii) NLXWs are not unique, in the sense that inifinitely
many solutions can be found for fixed $\beta$ and $\Gamma$.
This follows also from (ii) since also
linear paraxial X-waves are a continuous family,
e.g. parametrized by $\Delta$ in Eq.~(\ref{lin});
(iv) NLXWs and bullets exist for different sign of
$\beta$,  entailing opposite phase shifts in Kerr media,
and different constraints ($\beta \le -\delta k/2$ for
NLXWs, $\beta<\delta k/2$ for bullets) in Eqs.~(\ref{shg}).
\newline\indent
Once the existence of nonlinear LWPs supported by
normal GVD is established, one might wonder about
their importance and observability. In the linear regime,
both Durnin $J_0$ beams \cite{durnin87}
and nonmonochromatic X-shaped LWPs \cite{saari97}
can be observed only by means of experimental arrangements
that adapt the input to the LWP (e.g., by means of lensacons or axicons).
Vice\-versa, we have found that the interplay of the nonlinearity
and normal GVD is responsible for a  universal mechanism,
namely colored conical emission (CE), that allows
for the self-induced spectral (in $\Omega$-$K$) reshaping
necessary to turn conventional, e.g. gaussian, pulsed beams into X-waves.
In fact, it is known that in Kerr media \cite{liou92}
the (modulational) stability analysis of the cw
plane-wave solution $u=e^{i\Gamma \zeta}$ of Eq.~(1)
yields exponential amplification of conical plane-wave (or Bessel)
perturbations with wavevector $K$ and frequency detuning $\Omega$,
such to yield real values of gain
$g=\{[(D_1\Omega^2-K^2)^2 + \Gamma]^2-\Gamma^2\}^{1/2}$.
This analysis can be readily generalized for the
cw plane-wave eigensolutions $u_1=u_{10} \exp(-i \beta \zeta)$
$u_2=u_{20} \exp[-i (\beta+\delta k) \zeta]$ of Eqs.~(2-3) \cite{trillo95}
to obtain the gain $g=[b \pm (b^2-c)^{1/2} ]^{1/2}$,
where $2b = \Gamma(u_{20}^2-2u_{10}^2)-\Omega_1-\Omega_2$,
$c = (\Omega_1 \Omega_2-\Gamma u_{10}^2)^2 - \Gamma u_{20}^2 \Omega_2^2$,
and $\Omega_m \equiv \sigma_m K^2 - d_m \Omega^2 + m\beta - (m-1)\delta k$,
$m=1,2$.
In the normal GVD regime ($d_m>0$) these expressions entail CE, i.e.
preferential amplification of waves at $K$ (angles)
linearly increasing with frequency detuning.
By comparing (see Fig.~\ref{fig3}) the CE gain
with the NLXW spectrum, obtained by inverting
Eq.~(\ref{spectrum}), it is clear that the
instability provides amplification at $K-\Omega$ pairs
that favour the formation of NLXWs.
Although the stability analysis is carried
out for cw plane-wave pum\-ping beams,
CE occurs also from tightly-focused short-pulse
input beams \cite{liou92}.
In this case, we expect CE to amplify frequencies $K-\Omega$
contained in the broad input spectrum,
while preserving the phase coherence between
different spectral components necessary for
the formation of a NLXW.
\begin{figure}
\caption{Comparison of NLXW and CE gain in the spectral domain:
(a) $|S_1(K,\Omega)|$ transform of the field $u_1$ in
Fig.~2(a) ($|S_2(K,\Omega)|$ is similar);
(b) CE domains for the out-of-phase
cw plane-wave eigenmode with the same parameters.}
\label{fig3} \end{figure}
In order to support this conjecture and prove that NLXWs
are the key to understand the dynamics of experiments
carried out with narrow beams and short pulses in the normal GVD regime,
we have conducted numerical simulations of the propagation.
This is crucial also to assess the observability of such type
of LWPs with finite-energy, which are the nonlinear counterpart
of monochromatic finite-aperture $J_0$ \cite{durnin87}
or Bessel-Gauss beams,
and nonmonochromatic finite-energy linear X-waves \cite{lg92,saari97},
observable in the real world.
While extensive results will be reported elsewhere,
we focus specifically on quadratic media [Eqs.~(\ref{shg})] in the large
negative
mismatch limit, where the field $u_1$ plays a leading role,
and experiences an effective
focusing Kerr effect \cite{spatbook,liu99}.
This case has twofold relevance: (i) higher-order effects not
included in our mo\-dels (Raman, self-steepening, space-time coupling,
saturation, etc.)
have lesser impact on the propagation as compared with true Kerr media;
(ii) experimental results in Lithium Triborate (LBO)
indicates the occurence of pulse compression in spite
of the fact that the medium is normally dispersive \cite{cleo01}.
We model the latter case assuming $\chi_2 \simeq 7\times10^{-5}~{\rm W}^{1/2}$,
$k_1''=0.015$~ps$^2$/m, $k_2''=0.07$~ps$^2$/m,
a mismatch $\Delta k=-30~{\rm cm}^{-1}$ ($\delta k=-180$),
and a spatio-temporal input gaussian beam $u_1(Z=0)$ with
FWHM $170$ fsec duration and $65~\mu$m beam width.
Figure~\ref{fig4} (top panel) shows the output intensity in a 4 cm crystal,
in the GVM-matched case ($\lambda=1.3 \mu$m, $T=35$ C in LBO) for
an intensity $I_p=50~{\rm GW/cm}^2$ ($Z_{nl}=0.6$~mm).
As shown, while a moderate fraction of the energy lags
behind and in front in the form of pulse satellites,
the main portion of the beam develops the characteristic
structure of a NLXW.
Importantly, we observe that the formation of NLXW-type
of LWPs in Fig.~\ref{fig4} is accompanied by strong pulse compression.
In fact, it can be shown that the peculiar spatio-temporal
structure of a NLXW leads to an effective GVD,
which  we obtain in the form $k''_{eff} \equiv \frac{d^2 k_z}{d\omega^2}
\simeq \frac{d^2}{d\omega^2}[k(\omega)-\frac{k_t^2}{2k_0}] \simeq
k''-k_0(\frac{d\theta}{d\omega})^2$, where the
angle $\theta \simeq \sin \theta = k_t/k_0$,
and $k_t=(k_0^2-k_z^2)^{1/2}=K/(|\beta| W_0)$.
Therefore the dispersive contribution which stems from the angular dispersion
$\theta=\theta(\omega)$ counteracts the material (normal) GVD, leading to
an effective anomalous GVD, which in turn explains the compression.
\begin{figure}
\caption{Output intensity profile $|E_1|^2=I_p|u_1|^2$, as obtained from Eqs.
(2),  (see text for coefficient values) in the GVM-matched case $v=0$
(top panel), and with GVM, $v \neq 0$ (bottom panel).}
\label{fig4} \end{figure}
\newline\indent
NLXWs are robust also against strong GVM.
First, when $v \neq 0$ in Eqs.~(\ref{shg}), we find CE
as well as NLXW solutions of Eqs.~(\ref{shg})
characterized by an additional complex phase profile
(details will be given elsewhere).
On the other hand, our simulations show that,
at sufficiently high intensity,
after a short distance, the launched ($u_1$)
and generated ($u_2$) fields tend to develop
spontaneously NLXW shapes, meanwhile leading
to nonlinear walk-off compensation ($u_{1,2}$ travel
locked together).
For instance in Fig.~\ref{fig4} (bottom panel) we show the profile
$|u_1|^2$ obtained for an input intensity $I_p=70$ GW/cm$^2$ after $1.5$
mm propagation  in LBO at $\lambda=1.06 \mu$m,
where a GVM as large as $\delta V=45$ ps/m ($v=75$) is compensated.
\newline\indent
Our results show that, contrary to common belief,
nonlinear space-time localization takes place
also in normally dispersive media. NLXWs are the eigenmodes
of 1+3D paraxial wave propagation models, and will be central
to interpret correctly numerical and experimental results.
Having restricted ourselves to NLXWs that travel at the natural group-velocity
of light, further work will be devoted to search for
NLXW solutions with finite energy (of which we have given numerical evidence),
and/or sub- or super-luminal nature, as well as the role played by
nonparaxiality and higher-order nonlinear terms.
\newline\indent
Funds from MIUR (PRIN project) and INFM (PAIS project)
are gratefully acknowledged. G.V. thanks
Lithuanian Science and Studies Foundation (grant T-491). J.T. thanks
Secretaria de Estado y Universidades, Spain. C.C. thanks Fondazione Tronchetti
Provera.



\begin{references}

\bibitem{spatbook} S. Trillo and W.E. Torruellas, eds.,
{\em Spatial Solitons} (Springer, Berlin, 2001).

\bibitem{agrawal} G.P. Agrawal,
{\em Nonlinear fiber optics} (Academic Press, 1995).

\bibitem{ran96}  J.K. Ranka, R.W. Schirmer, and A.L. Gaeta
Phys. Rev. Lett. {\bf 77}, 3783 (1996).

\bibitem{zoz99} A. A. Zozulya,
S.A. Diddams, A.G. Van Engen, and T.S. Clement,
Phys. Rev. Lett. {\bf 82}, 1430 (1999).

\bibitem{liu99} X. Liu, L.J. Qian, and F.W. Wise,
Phys. Rev. Lett. {\bf 82}, 4631 (1999).

\bibitem{kopr00} I.G. Koprinkov {\it et al.},
Phys. Rev. Lett. {\bf 84}, 3847 (2000);
A.L. Gaeta and F. Wise,
Phys. Rev. Lett. {\bf 87}, 229401 (2001).

\bibitem{eis01} H.S. Eisenberg {\it et al.},
Phys. Rev. Lett. {\bf 87}, 043902 (2001).

\bibitem{tzo01} S. Tzortzakis {\it et al.},
Phys. Rev. Lett. {\bf 87}, 213902 (2001).

\bibitem{silb90} Y. Silberberg,
Opt. Lett. {\bf 15}, 1282 (1990).

\bibitem{acoutheory}
P.R. Stepanishen and J. Sun,
J. Acoust. Soc. Am. {\bf 102}, 3308 (1997);
J. Salo, J. Fagerholm, A.T. Friberg, and M.M. Salomaa,
Phys. Rev. Lett. {\bf 83}, 1171 (1999).

\bibitem{opttheory} J. Salo, J. Fagerholm, A.T. Friberg, and M.M. Salomaa,
Phys. Rev. E {\bf 62}, 4261-4275 (2000);
K. Reivelt and P. Saari,
J. Opt. Soc. Am. A {\bf 17}, 1785 (2000).

\bibitem{durnin87} J. Durnin, J.J. Miceli, and J.H. Eberly,
Phys. Rev. Lett. {\bf 58}, 1499 (1987);

\bibitem{lg92}
J. Lu and J.F. Greenleaf,
IEEE Trans. Ultrason. Ferrelec. Freq. contr. {\bf 39},
441-446 (1992).

\bibitem{saari97} P. Saari and K. Reivelt,
Phys. Rev. Lett. {\bf 79}, 4135 (1997);
H. S\"{o}najalg, M. Rtsep, and P. Saari,
Opt. Lett. {\bf 22}, 310 (1997).

\bibitem{chqz} C. Canuto {\it et al.},
{\em Spectral methods in fluidodynamics} (Springer, New York, 1988).

\bibitem{bessel}
P. Di Trapani {\it et al.},
Phys. Rev. Lett. {\bf 81}, 5133 (1998).

\bibitem{liou92} L.W. Liou, X.D. Cao, C.J. McKinstrie,
and G.P. Agrawal,
Phys. Rev. A {\bf 46} 4202-4208 (1992);
G.G. Luther, A.C. Newell, J.V. Moloney, and E.M. Wright,
Opt. Lett. {\bf 19} 789-791 (1994).

\bibitem{trillo95} S. Trillo and P. Ferro,
Opt. Lett. {\bf 20}, 438 (1995).

\bibitem{cleo01} G. Valiulis {\it et al.},
Quantum Electronics and Laser Science (QELS 2001)
Conference, (Optical Society of America, Washington DC, 2001),
postdealine paper QPD10-1.

\bibitem{recami} M. Zamboni-Rached, E. Recami, and H.E.
Hernandez-Figueroa, arXic:physics/0109062 v2 (2001).

\end{references}
\end{document}